\documentclass[%
 reprint,
 amsmath,amssymb,
 aps,
]{revtex4-2}

\usepackage{lipsum}
\usepackage{graphicx}
\usepackage{dcolumn}
\usepackage{bm}
\usepackage{color}
\usepackage{comment}

\usepackage{braket}
\begin{document}
\preprint{APS/123-QED}
\title{Topological phases and entanglement in real space for 1D SSH topological insulator: effects of first and second neighbor-hoppings}
\author{Leonardo A. Navarro-Labastida$^{1,2,*}$, Francisco A. Dom\'inguez-Serna$^3$ and F. Rojas$^{2,}$}
\date{March 2022}
\email{leonardo.navarro@estudiantes.fisica.unam.mx\\
frojas@cnyn.unam.mx}
\affiliation{%
$^1$Depto. de Sistemas Complejos, Instituto de F\'isica, \\ Universidad Nacional Aut\'onoma de M\'exico (UNAM)\\
Apdo. Postal 20-364, 01000, CDMX, M\'exico.\\
$^2$Depto. de F\'isica , Centro de Nanociencias y Nanotecnolog\'ia, Universidad Nacional Aut\'onoma de M\'exico (UNAM), Apdo. Postal 22800, Ensenada, Baja California, M\'exico.\\
$^3$C\'atedras CONACYT, Centro de Investigaci\'on Cient\'ifica y de Educaci\'on Superior de Ensenada,
Apartado Postal 2732, BC 22860 Ensenada, M\'exico.}%
\begin{abstract}
The hybrid atoms-cell site entanglement in a one-dimensional Su-Schrieffer-Heeger (SSH) topological insulator with first and second neighbor hopping in space representation of finite chains is analyzed. The geometrical phase is calculated by the Resta electric polarization and the entanglement in the atomic basis by the Schmidt number. A relation between entanglement and the topological phase transitions (TPT) is given since the Schmidt number has local critical points of maximal entangled (ME) states in the singularities of the geometrical phase. States with second-neighbors have higher entanglement than first-neighbors hopping. The general conditions to produce ME hybrid Bell states and the localization-entanglement relation are given.
\end{abstract}
\maketitle 
\section{Introduction} 
Topological systems promise to be materials with various implementations \cite{Hasan2010, Xiao2011}; in condensed matter physics. These materials gained interest in recent years due to their peculiar properties like efficient transport in electronic hetero-structures \cite{HANG2021}, high thermal conductivity \cite{Platero2018}, favorable mechanical properties under strains \cite{Kane2014, Naumis2014}, minimization of thermal noise \cite{WEBER2016}, and decoherence effects in open systems \cite{SHAOLIN2020}. \\
The property of topological robustness protects these materials from quantum fluctuations or defects in the system \cite{Guinea2009} and presents protected states, also known as zero-mode energy states \cite{Xiao2011}.\\
There are already some works that mention peculiarities of these types of materials \cite{LONG2020,Taboada2015, Xiao2011, SHAOLIN2020, JAY2010, Navarro2021}, and in general, due to the robustness of these materials, there are more quantum correlations which allow a greater degree of efficiency in electronic transport \cite{YIXIN2010,Hauke2014} related with the appearance of flat bands. Therefore, topological materials are expected to be suitable for quantum information processing \cite{NIU2010,JAY2010,Jun2014,YU2006}.\\
One of the topics in quantum information theory is the study of entanglement and quantum correlations involved in condensed matter systems \cite{Ping2019, Jaeyoon2017,2021TakamiT}. The origin of the relationship between quantum correlation metrics and geometric phases comes from the Fubini-Study geometrical quantum tensor of the complex projective space in the projective geometry of Hilbert space \cite{Brody2001}. The connection between topological materials in condensed matter physics and manipulating qubits with entanglement properties in quantum information theory \cite{Horodecki2009} opens a new research area to create new technology, like, topological quantum computing, cryptography, and quantum security \cite{VALERIO2009, JIAN2012, NICOLAS2012}.\\
There are several experimental setups \cite{Ozawa2020,YANG2019,2021Oliveira,2020YANGG,2018Yuting}, that studied some properties as anomalous transport, decoherence times, and thermal capacity.
More recently is broadly studied photonics systems and the detection of topological states in light-matter devices like the SSH model \cite{Schrieffer1980} crystal photonic systems promise to have robust transport due to the presence of these protected gapless states \cite{Ozawa2020}. \\
In the SSH model, modulation of the hoppings generates a phase transition between a metal-insulator behavior due to the Piers instability of the deformation of hoppings. 
In this work, we presented a characterization of the TPT via the Schmidt number metric \cite{Knight2006, Sperling2011} as a measure of entanglement in the simple \cite{Obana2019,Kuno2019} and extended \cite{CHAO2010} SSH models. Also, the relation entanglement-localization and topology are discussed.  

\section{1D SSH topological insulator}
The system of study is the SSH Hamiltonian \cite{Schrieffer1980}, which is a tight-binding model of a wire with alternating single and double hopping (Fig. 1). The basis of the wire is constructed by a cell of two types of atoms A and B.
\begin{figure}[h!]
\includegraphics[width=8.2cm]{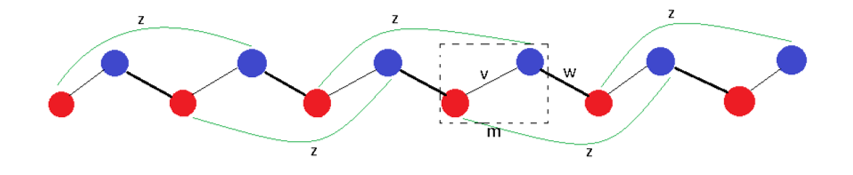}
\caption{Extended SSH model. Topological wire, with  intra $v$ (thin lines) and inter $w$ (thick lines) hopping and an additional second neighborhood hopping $z$ (green lines). The index $m$ (tick lines) indicates the cell number and atoms A (Red) and B (Blue), with $N=6$ unit cells.}
\end{figure}\\

The Hamiltonian in real space can be written as, 
\begin{equation}
\begin{split}
\hat{H}_{e}&=\hat{H}_{s}+z\sum^N_{m}(\ket{m+1}\bra{m}\otimes\ket{B}\bra{A}+h.c.)\\
 &= \hat{H}_{s}+z\sum^N_{m}(\ket{m+1}\bra{m}\otimes\hat{\sigma}_{-}+h.c.)
\end{split},
\label{Ham}
\end{equation}
where $z$ is the second neighbor hopping  and $\hat{H}_{s}=\sum^N_{m}(v\ket{m}\bra{m}\otimes\ket{A}\bra{B}+w\ket{m+1}\bra{m}\otimes\ket{A}\bra{B}) + h.c.
=v\sum^N_{m}\ket{m}\bra{m}\otimes\hat{\sigma}_{x}+w\sum^N_{m}(\ket{m+1}\bra{m}\otimes\hat{\sigma}_{+}+ h.c)$ is the simple SSH Hamiltonian where $\hat{\sigma}_{i}$ with $i=x,y,z$ are the Pauli matrices with $\hat{\sigma}_{+}=\frac{\hat{\sigma}_{x}+i\hat{\sigma}_{y}}{2}$ and $\hat{\sigma}_{-}=\frac{\hat{\sigma}_{x}-i\hat{\sigma}_{y}}{2}$. Here "$v$" and "$w$" represents intra-cell and inter-cell hopping in 1D wire, $\ket{m}$ is the index of the cell number, $\ket{A}$ and $\ket{B}$ are the occupation type atom in a cell.
This model defines the basic form for the bipartite system of $2\otimes N$ dimension. \\
The eigenfunctions can be expanded as a superposition of the composite states, $\ket{m,\alpha}= \ket{m} \otimes\ket{\alpha}\in H_{ext}\otimes H_{int}$, here $H_{ext}\rightarrow \ket{m}$ is the external dimension or degree of freedom related to the position over the chain, while $H_{int}\rightarrow \ket{\alpha}$ represent the internal degree of freedom with $\alpha\in{(A,B)}$ related to the type of atom. \\
The eigensolution of the Hamiltonian (Eq. \ref{Ham}), $H\ket{\psi_n}=E_{n} \ket{\psi_n}$ can be expressed as a combination of the composite state, $\ket{\psi_n}=\sum^N_{m}(C^A_{m,n}\ket{m,A}+C^B_{m,n}\ket{m,B})$,  where $C^{\alpha}_{m}$ are the amplitude of probability of the particle to be in cell $\ket{m} $ and atom $\ket{\alpha}$. 
\section{Geometrical phases and electrical polarization}
To understand geometrical phases in our models, we characterized the phases in the k-space, and we provided a specific procedure to determine it in real space based on the calculation of electric polarization in periodic systems introduced by Resta \cite{RESTA2000}. \\
The Hamiltonian in real space can reduce to a $2\times2$ matrix in k-space as $H=\sum_k\psi^{\dagger}_kH(k)\psi_k$, where $\psi_k=(a_k, b_k)^T$ are the Nambu spinor. For the extended Hamiltonian we get, 
\begin{equation}
\begin{split}
H(k)=\begin{pmatrix}
0 & v+we^{-ik}+ze^{ik} \\
v+we^{ik}+ze^{-ik} & 0
\end{pmatrix},
\end{split}
\end{equation}
which can be rewritten in terms of Pauli matrices as $H(k)=\bm{h(k)}\cdot \bm{\sigma}$, where $h(k)=(v+(w+z)\cos{k},(w-z)\sin{k},0)$ is a vector in the plane $h_x-h_y$ that maps an ellipsoid centered on $v$. Also in complex plane is $h(k)=h_x(k)+ih_y(k)$. \\

The energy in k-space is given by $\epsilon^{\pm}_k=\pm\sqrt{v^2+w^2+z^2+2v(w+z)\cos{k}+2wz\cos{2k}}$ with  associated eigenvectors $\ket{u^{\pm}_k}=\frac{1}{\sqrt{2}}(\pm e^{-i\phi(k)},1)$ where $\phi(k)=\tan^{-1}{\frac{h_y}{h_x}}$. Using $z=0$ the simple SSH case can be recovered with  $h(k)=(v+w\cos{k},w\sin{k})$ being now a circle center in $v$. Clearly, the energy spectrum is quiral and therefore we have that $E_{-}=-E_{+}$ and also have the property of chirality of the Hamiltonian $\sigma_z H\sigma_z=-H$, or $\lbrace \sigma_z, H \rbrace$=0 with $\sigma^2_z=1$. 
\\
The topology of the 1D SSH model is characterized by the winding number that is related to the Berry phase or geometric phase for an adiabatic system \cite{BARRY1983}. The winding number for the extended Hamiltonian can be written as, 
\begin{equation}
\begin{split}
\zeta & =\frac{1}{2\pi i}\int_{C}\frac{d}{dk}ln{(h(k))}\\
 & =\frac{1}{2\pi i}\int_{C}\frac{d}{dk}ln{(v+we^{ik}+ze^{-ik})},\\
\end{split}
\end{equation}
where $C$ is the Brillouin zone $k\in [-\pi,\pi]$. We get the simple SSH model setting $z=0$ in the integral (Eq. 3); therefore, the winding number has values, 
\begin{equation}
\begin{split}
\zeta & =\begin{cases}
      0, &  v > w\\
      1, &  v < w 
    \end{cases}    ,
\end{split}
\end{equation}
where the topological region (TR) is $v>w$, the trivial region $v<w$ and the singularity of the winding number occurs in $v=w$. 

For the extended SSH model $z\neq0$, we obtain the winding number,
\begin{equation}
\begin{split}
\zeta & =\begin{cases}
      0, &  v > w+z\\
      1, &  v < w+z , w > z\\
      -1, & v < w+z , w < z 
    \end{cases}    ,
\end{split}
\label{r1}
\end{equation}
when the TR happens in $v<w+z$ and trivial region in $v>w+z$.\\

For $\zeta=\pm 1$, we have a topological insulator with the sign related to the direction of the path over the curve. This topological invariant is related to the Berry phase as $\gamma=\pi\zeta$ and also with the electrical polarization $P=e\frac{\zeta}{2}$. \\ 
In the extended SSH model there are two points of TPT $v=w+z$ and $w=z$, the first one $v=w+z$ is related to TPT between trivial and non-trivial regions $\zeta=0\rightarrow 1$ while $w=z$ is related to TPT between TR $\zeta=\pm1\rightarrow \mp1$.
\\

However, geometrical phases are determined in k-space, where the system has periodic boundary conditions. Therefore we need to use another more convenient procedure in real space where the non-separable property of the bipartite system still remains. For this reason, we calculate the electrical polarization by the definition of Resta polarization \cite{RESTA2000, RESTA1998},

\begin{equation}
\begin{split}
P_n=\frac{e}{2\pi}Im\ln{[\bra{\psi_n}e^{i\delta\hat{x}}\ket{\psi_n}]} 
\end{split}
\end{equation}
where $\delta=\frac{2\pi}{Na}$ and $\hat{x}=\sum^N_{m}\hat{x}_m = \sum^N_{m}m[\ket{m,A}+\ket{m,B}]$, with the charge of electron "$e$" and the atomic distance "$a$" in natural unities $e=a=1$ and $\hat{X}=e^{i\delta\hat{x}}$. Its follows that geometrical phase is $\gamma_n=Im\ln{[\bra{\psi_n}e^{i\delta\hat{x}}\ket{\psi_n}]}$, so $P_n=\frac{e}{2\pi}\gamma_n$, electrical polarization is proportional to the geometrical phase \cite{MARTIN1994}. The idea in this formulation is consider a new operator of position for a composite system with periodic boundary conditions in real space \cite{MICHAEL2017}. For the SSH basis wavefunction can write the electric polarization \cite{balazss2021} as $P_n=\frac{1}{2\pi}Im\ln{[\sum^N_{m}e^{i\delta m}(|C^A_{m,n}|^2+|C^B_{m,n}|^2)]}$. 

\section{Schmidt Number and entanglement}
For the measure of entanglement use the Schmidt number \cite{Bagdanov2007,Eberley2006} because is defined as the metric of entanglement in pure bipartite systems and can be described by bi-orthonormal wavefunctions as $\ket{\psi^{ext,int}}=\sum^k_{n=1}\sqrt{p_n}\ket{u^{ext}_n, w^{int}_n}$ \cite{Vogel2006}, where $k\leq dim[min(ext,int)]$ and $\mathcal{H}=\mathcal{H}^{ext}\otimes\mathcal{H}^{int}$, which allows to characterize the degree of entanglement through the Schmidt number $K$, defined as follows
\begin{equation}
\begin{split}
K = \frac{1}{\sum_{i} \lambda^2_i} = \frac{1}{Tr(\rho^2_r)}
\end{split}
\end{equation}\\
where $\rho_r$ is the density matrix of the reduced space $\rho_r=Tr_p(\rho^{ext,int})$  For non-separable system its follows that $\rho^{ext,int}\neq \rho^{ext}\otimes\rho^{int}$  with the property $Tr(\rho^{ext})=Tr(\rho^{int})=1$. 
\\
The Schmidt number is also defined as the metric of entanglement for the SSH model in the reduced space of the qubit formed by the two-level system referred to as the type of atom $A$ or $B$. The total dimension of the SSH model is $2\otimes N$ with $N$ the number of sites over the chain. By doing the partial trace, we have that the reduced matrix is a $2\times 2$ matrix. \\
We considered that the hoppings are real numbers, therefore, the wave function of the eigenstate $n$ is $\ket{\psi_n}=\sum^N_{m}(C^A_{m,n}\ket{m,A}+C^B_{m,n}\ket{m,B})=\frac{1}{\sqrt{2}}(\ket{\phi^A}\otimes\ket{A}+\ket{\phi^B}\otimes\ket{B})$, where $\ket{\phi^{\alpha}}=\sqrt{2}\sum^N_{m}C^{\alpha}_{m,n}\ket{m}$, expressing in this form $\ket{\psi_n}$ is clear that has a non-separable basis, where $\ket{\phi^{A}}$ and $\ket{\phi^{B}}$ are mutual orthogonal. 
\\
The Schmidt number can be interpreted from a geometric point of view when the reduced density matrix represents a two dimensional system, in this condition the density matrix defines a Bloch vector of the form
$\langle \bm{r}\rangle=\langle\bm{\sigma} \rangle=Tr(\rho_n\bm{\sigma})$, where $\rho_n=\ket{\psi_n}\bra{\psi_n}$ is the density matrix of the pure state $\ket{\psi_n}$ and $\bm{\sigma}=(\sigma_x,\sigma_y,\sigma_z)$ are the Pauli matrices. For a qubit system $\rho=\frac{1}{2}(\mathbb{I}+ \bm{r}\cdot\bm{\sigma})$, using this notation Schmidt number can be rewritten as, 
$K=\frac{d}{1+|\bm{r}|^2}$, where $d$ is the dimension of the reduced density matrix in qubit system. \\

Here we consider $d=2$, due to the internal dimension of the basis atom $\ket{\alpha}$. The ME is reached when $K=2$, this occurs for Bloch vectors where $|\bm{r}| \rightarrow 0$, on the other hand, the system becomes separable when $K=1$ such that the Bloch vector is $|\bm{r}| \rightarrow 1$. The ME states are of the hybrid Bell states for the structure of the wavefunction. \\
The hybrid entanglement is between cell sites and atoms, these must satisfy conditions; i) $ \braket{\phi^{A}|\phi^{A}}=\sum_{m}|C^A_m|^2=\frac{1}{2}$,  ii) $\braket{\phi^{B}|\phi^{B}}=\sum_{m}|C^B_m|^2=\frac{1}{2}$ and iii) $\braket{\phi^{A}|\phi^{B}}=\sum_{m}(C^A_m)^{*}C^B_m=0$. Using this basis we can write Schmidt number as, $K_n=[(\sum_{m}|C^A_{m,n}|^2)^2+(\sum_{m}|C^B_{m,n}|^2)^2+2(\sum_{m}C^A_{m,n} C^B_{m,n})^2]^{-1}$. 
\\
The normalization for $\psi$ leads to the obvious condition for the coefficients $\sum_{m,n}|C_{mn}|^2=1$, which shows that each squared coefficient can be interpreted as a weight (probability). The average probability $|C_{mn}|^2$ is then given by $\sum_{m,n}|C_{mn}|^4$. The inverse of this is the ‘number’ of effectively non-zero probabilities \cite{Sperling2011}, so a degree of correlation $K$ is defined in this way, $K=1/\sum_{m,n}|C_{mn}|^4$. \\
In contrast to some other similar and also ‘natural‘ definitions, K has the following desirable properties: (a) it is independent of the representation of the wavefunction, so that, for example, K is the same in configuration and momentum space; (b) K is also gauge invariant, which is important for systems in the presence of electromagnetic (laser) fields; and (c) it obviously achieves its minimum value of 1 for the least correlated state. 
\\
An important case to analyze for entanglement aspects is the dimerized limit, the bulk in the fully dimerized limits has flat bands. These consist of even energy $E = +1$ and odd
energy $E=-1$ superposition of two sites forming a dimer. Trivial dimerized case occurs for $v=1$ and $w=0$, so, the eigenvalue equation reduce to $\hat{H}(\ket{m,A}\pm\ket{m,B})=\pm(\ket{m,A}\pm\ket{m,B})$. The topological dimerized case occurs for $v=0$ and $w=1$, therefore, the eigenvalue equation reduce to $\hat{H}(\ket{m,B}\pm\ket{m+1,A})=\pm(\ket{m,B}\pm\ket{m+1,A})$. For topological dimerized limit, the corresponding normalize states are $\psi^{max}_{\pm}=\frac{1}{\sqrt{2}}[\ket{m,B}\pm\ket{m+1,A}]$ and the reduced density matrix is 
\begin{equation}
\begin{split}
\rho^{max}_{r,\pm}=\frac{1}{2}[\delta_{m,m^{\prime}}\ket{B}\bra{B}+\delta_{m+1,m^{\prime}+1}\ket{A}\bra{A}]
\end{split},
\end{equation}
these states has a Schmidt number,
\begin{equation}
\begin{split}
K^{max}=\frac{1}{Tr[(\rho^{max}_{\pm})^2]}=2 
\end{split},
\label{r2}
\end{equation}
therefore, in the topological dimerized limit, there is an entangled state with a maximum Schmidt number. For other hand, trivial dimerized limit have normalize states as $\psi^{min}_{\pm}=\frac{1}{\sqrt{2}}[\ket{m,A}\pm\ket{m,B}]$ and the reduced density matrix is
\begin{equation}
\begin{split}
\rho^{min}_{r, \pm}&=\frac{1}{2}[\delta_{m,m^{\prime}}\ket{B}\bra{B}+\delta_{m,m^{\prime}}\ket{A}\bra{A}\\ &\pm\delta_{m,m^{\prime}}\ket{B}\bra{A}\pm\delta_{m,m^{\prime}}\ket{A}\bra{B}]
\end{split},
\end{equation}
and the corresponding Schmidt number is
\begin{equation}
\begin{split}
K^{min}=\frac{1}{Tr[(\rho^{min}_{\pm})^2]}=1
\end{split}.
\end{equation}
Therefore, in the trivial dimerized limit, the system becomes disentangled. In both fully dimerized limits, the energy eigenvalues are independent of the wavenumber, $E(k) = 1$. In this so-called flat-band limit, the group velocity is zero, which again shows that as the chain falls apart to dimers, a particle input into the bulk will not spread along the chain. In k-space trivial and topological dimerized limits are $\hat{H}(k)=\hat{\sigma}_x$ and $\hat{H}(k)=\hat{\sigma}_x\cos{k}+\hat{\sigma}_y\sin{k}$.

\section{Results}
\subsection{Simple SSH $z=0$}
\begin{figure}[ht!]
    \includegraphics[width=8.65cm]{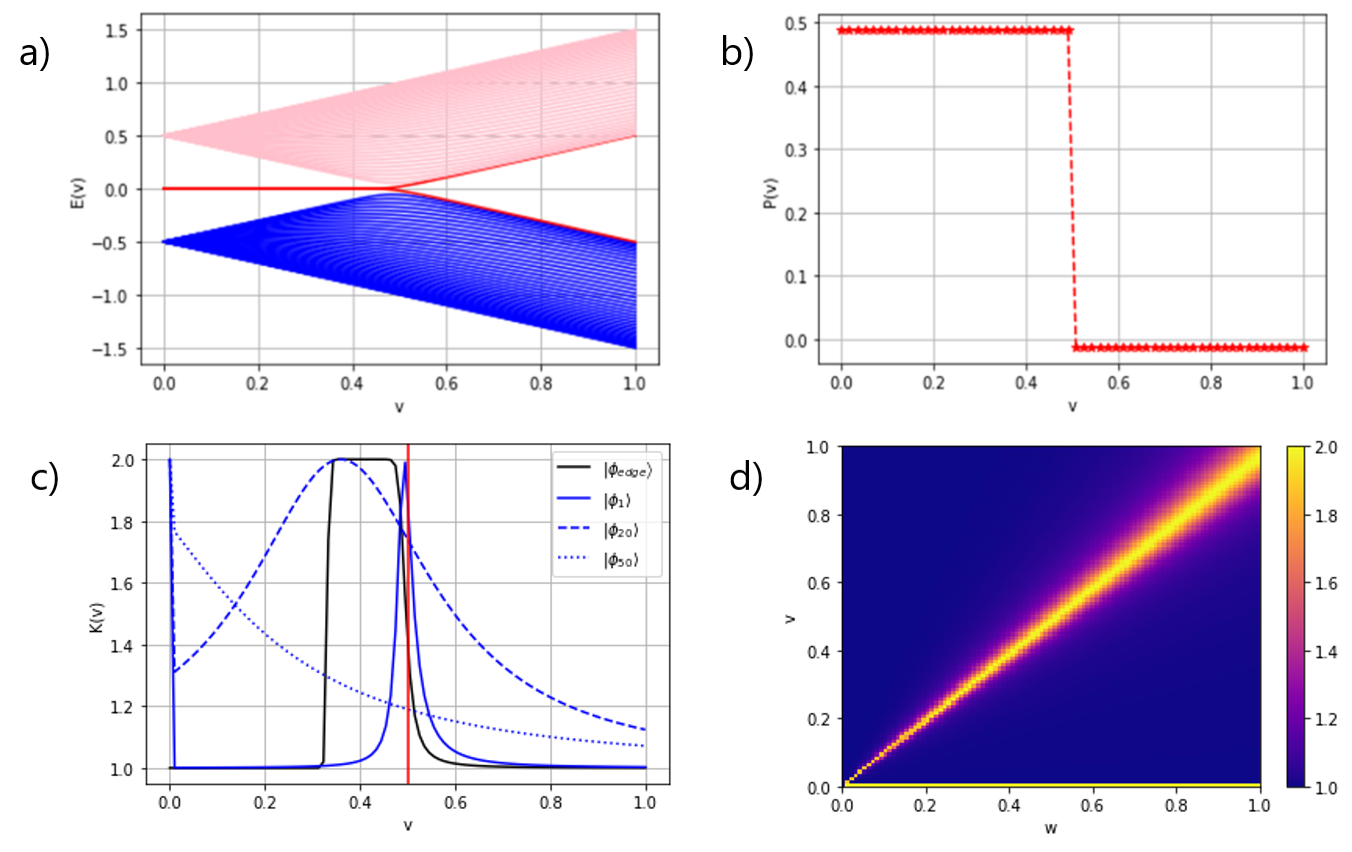}
    \caption{(a) Energy spectrum in the simple SSH model for $N=40$ unit cells and the inter-hopping is $w=0.5$ and the second neighbor hopping $z=0$. (b) Resta polarization for the quantum state $\ket{\psi_1}$. (c) Schmidt Number as a function of the intra-hopping parameter $K(v)$, the inter-hopping is $w=0.5$. The red vertical line specifies the transition point between the trivial $(v<w)$ and non-trivial $(w>v)$ zones. The Schmidt number for the bulk quantum states $\ket{\psi_1}$, $\ket{\psi_{20}}$, $\ket{\psi_{50}}$ and $\ket{\psi_{edge}}$ related to one of the edge state are plotted. (d) we see that in the TPT point $v=w$ the quantum state $\ket{\psi_1}$ related to the nearest zero energy is ME.}
    \label{ZCEROO}
\end{figure}

For the simple SSH model, two kinds of eigenstates can be distinguished: i) edge states, in the region $v\leq w$ with the property of zero energy, $E=0$ and ii) bulk states, in the region $w\leq v$, these becomes exponentially and the energy gap increase (Fig. \ref{ZCEROO}(a)). Using the Resta definition for the electric polarization for the nearest states to the edge state $\ket{\psi_1}$ is shown in Fig. \ref{ZCEROO}(b).

The SSH model has a spectrum of $2N$ eigenvalues $\lbrace -E_N,...,-E_1,-E_0, E_0, E_1,..., E_N \rbrace$, where $\pm E_0$ would be the edge state in the topological regimen. To get a general picture and observe the behavior of each state, we study the bulk states$\ket{\psi_1}$, $\ket{\psi_{20}}$, $\ket{\psi_{50}}$ and one of the edge states $\ket{\psi_{edge}}$, for the case with $N=80$ unit cells. 

We observe that the TPT occurs in the singular point $v=w$ according to the geometrical phase definition Eq.4. The eigenstate $\psi_{1}$ characterizes the relationship between the quantum entanglement and topology just in the TPT.

The Schmidt number as function of $v$ is plotted in Fig. \ref{ZCEROO}(c) for fixed $w=0.5$. The red vertical line indicates the singularity point $v=w=0.5$, and the curves correspond to eigenstates $\ket{\psi_{edge}}$ (black line), $\ket{\psi_{1}}$ (blue solid line), $\ket{\psi_{20}}$ (blue dashed line) and $\ket{\psi_{50}}$ (blue pointed line). In the simple SSH Hamiltonian case, we see for the quantum states a direct relation between entanglement and geometrical phases in the TPT $v=w$.\\
The edge state is the most robust in the region $v<w$ with a maximal Schmidt number $K=2$. For the state $\ket{\psi_1}$ in the singular point $v=w$ of the winding number have a ME state. $\ket{\psi_1}$ is the nearest state to the edge state and has a narrow peak of a ME state. For the higher energy eigenstates,$\ket{\psi_{20}}$ and $\ket{\psi_{50}}$, the peak is broadened and shifted to small values for $v$. The case $v=0$ calculated in Eq. 9 which is the topological dimerized limit has ME.\\
In Fig. \ref{ZCEROO}(d) is shown the entanglement diagram of eigenstate $\ket{\psi_1}$ as a function of $v$ and $w$. The state $\ket{\psi_1}$ for all the values that satisfy the critical point condition $v=w$ has maximum Schmidt number and has a peak with as mall broaden. Away from this region, the system becomes disentangled.\\ 
Therefore, $\ket{\psi_1}$ is maximally localized as $\delta(v-w)$. In the region $v=w$, the Schmidt number is ME, but an increment in hoppings $v$ and $w$ produce that the localization of wavefunction becomes to broaden; however, always center in $v=w$ as a Gaussian distribution.

\subsection{Extended SSH $z\neq 0$}
\begin{figure}[h!]
    \includegraphics[width=8.6cm]{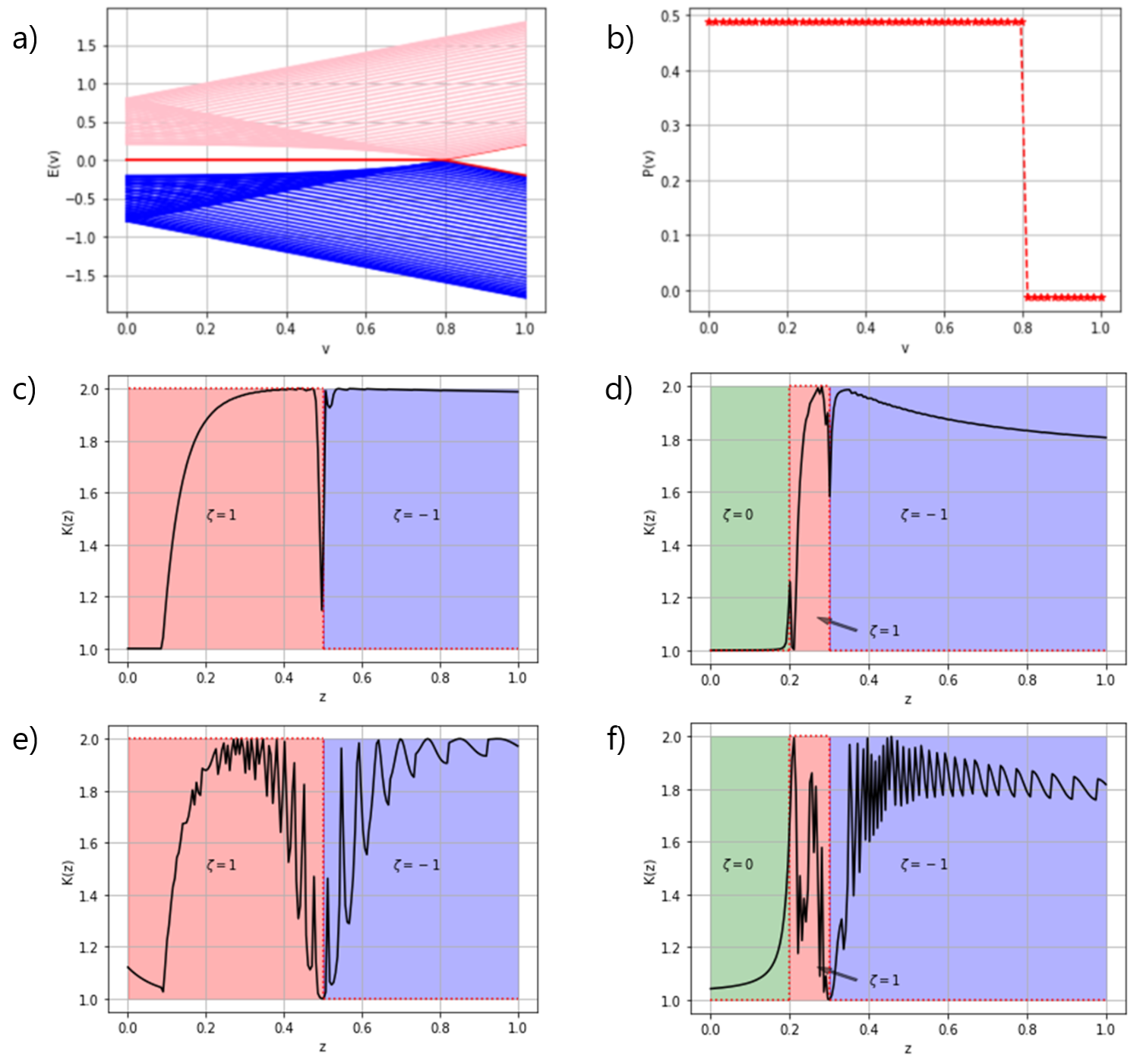}
    \caption{(a) Energy spectrum in the extended SSH model as function of $v$ for $N=40$ unit cells and $w=0.5$ and $z=0.3$. (b) Resta polarization for the quantum state $\ket{\psi_{1}}$. Schmidt number in extended SSH as function of $z$ for $N=300$ unit cell, for $\ket{\psi_1}$ with (c) $v=0.3$, $w=0.5$ and (d) $v=0.5$, $w=0.3$. For $\ket{\psi_{20}}$ with (e) $v=0.3$, $w=0.5$ and (f) $v=0.5$, $w=0.3$. }
    \label{S_ext}
\end{figure}

For the extended SSH model, the energy spectrum is similar to the simple case; however, edge states appear in the region $v<w+z$ (Fig. \ref{S_ext}(a)).

The Fig. \ref{S_ext}(b) shows the electric polarization for the nearest states to the edge state $\ket{\psi_1}$, non-trivial geometrical phase occurs in the region $v<w+z$. From Eq. \ref{r1}, this has three distinct values for the geometrical phase. For $\zeta=0$ is a trivial insulator and therefore the electric polarization $P_1=0$, while, for the topological cases $\zeta=\pm 1$ has $P_1=\pm\frac{e}{2}$.

In Fig. \ref{S_ext}(c) for $\ket{\psi_1}$, the Schmidt number has a TPT $\zeta=1\rightarrow -1$ when $w=z$ for $v=0.3$ and $w=0.5$. Also the Schmidt number as function of $z$ is characterized by the instantaneous lost of the ME. It is important to remark that in the region $w>z$ (red) reaches ME $K_1=2$ when $v<w+z$, and in the region $w<z$ (blue) always have ME states. 

On the other hand, in Fig. \ref{S_ext}(d) for $v=0.5$ and $w=0.3$, three regions can be distinguished. The region $v>w+z$ (green) has $\zeta=0$ and $K_1=1$, therefore, the state is separable. For $v<w+z$ and $w>z$ (red), $\zeta=1$ and $\ket{\psi_1}$ tends to a ME. In the region $v<w+z$ and $w<z$ (blue) $\zeta=-1$ also has a ME behavior. \\
Note that the TPT points $v=w+z$ and $w=z$ represent a different kind of degeneracy. This difference is visible looking at the Schmidt number. The point $v=w+z$ is ME singularity, and the state tends to reach hybrid Bell state, while $w=z$ is a disentanglement point, and the system becomes separable.
For higher energy states like $\ket{\psi_{20}}$, Schmidt number has more energy fluctuations in both topological phases $\zeta=\pm1$ (Fig. \ref{S_ext}(e)-(f)), therefore, topological protection does not prevent quantum fluctuations in higher energy levels. Note however that $\ket{\psi_{20}}$ also characterizes the critical points $v=w+z$ and $w=z$. 

\begin{figure}[h!]
    \includegraphics[width=8.5cm]{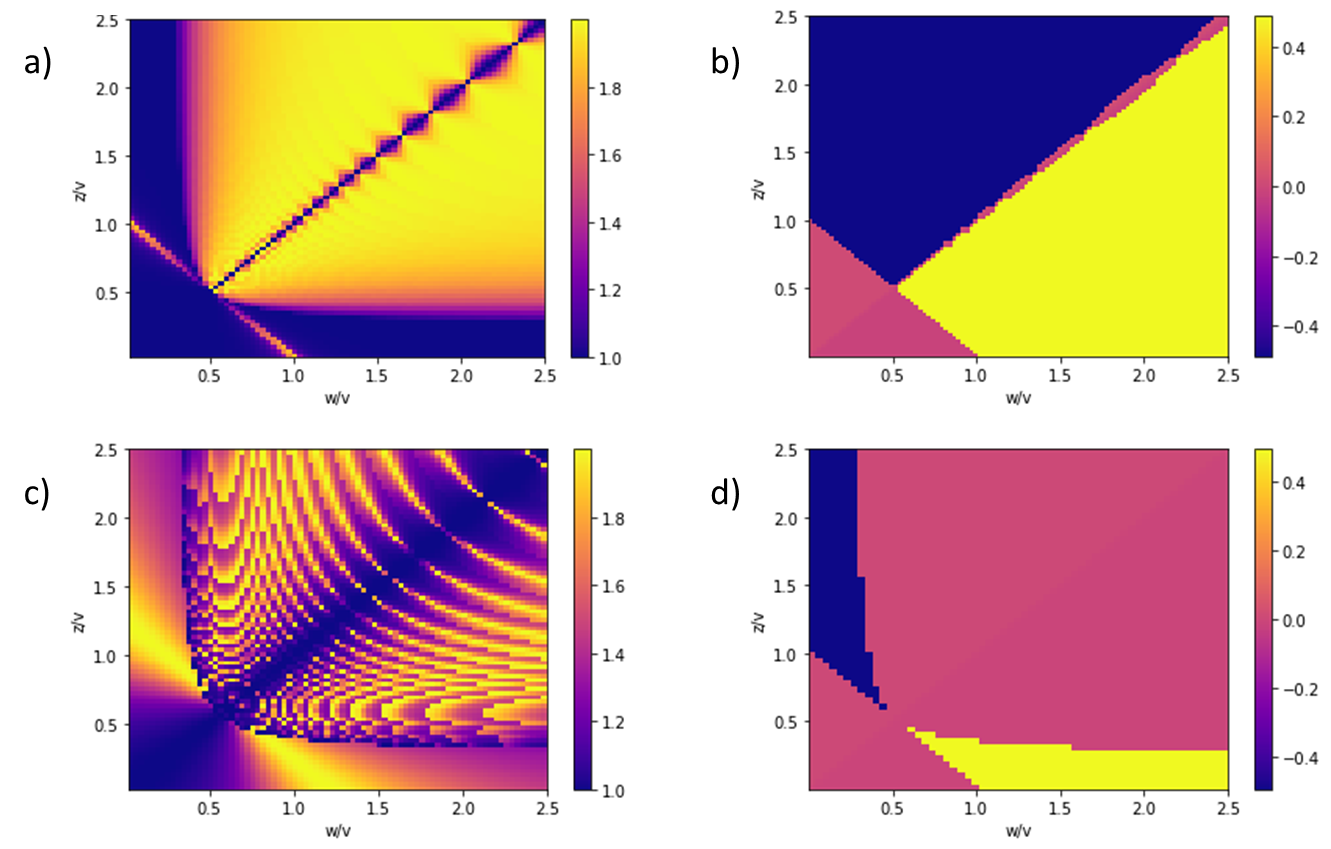}
    \caption{Schmidt number and electric polarization diagrams in the extended SSH model as function of $w$ and $z$ for $v=0.4$ and $N=300$ unit cells. For $\ket{\psi_1}$ (a) $K_1(\frac{w}{v},\frac{z}{v})$ and (b) $P_1(\frac{w}{v},\frac{z}{v})$. For $\ket{\psi_{20}}$ (c) $K_{20}(\frac{w}{v},\frac{z}{v})$ and (d) $P_{20}(\frac{w}{v},\frac{z}{v})$. }
    \label{SPz}
\end{figure}
To obtain a general picture of the whole parameters space influence on entanglement for the extended SSH model the diagram of the electric polarization $P(\frac{w}{v},\frac{z}{v})$ and the Schmidt number $K(\frac{w}{v},\frac{z}{v})$ with an intra-hopping of $v=0.4$ is shown in fig. \ref{SPz} (a)-(d).

For $\ket{\psi_{1}}$, the presence of topological phases indicates states with ME; however, note that not the entire region $v<w+z$ presents states with ME. The Schmidt number presents disentangled states in $w=z$ and entangled states in $v=w+z$ just in singularities of the winding number, which are also related to the transition feature of the topological insulator and with the broken chiral symmetry.
For $\frac{z}{v}<1$ there is a trivial electric polarization $P_1=0$ while $\frac{z}{v}>1$ has a non-trivial electric polarization $P_1=-\frac{1}{2}$.

When $z=0$, we recover the simple SSH model, and the critical transition point remains for $w=v$. The entanglement behavior for both topological phases $\zeta=\pm1$ is the same, and between there is a singular behavior that generates disentanglement. For eigenstate, $\ket{\psi_{20}}$, the singularities of winding number are still preserved. 
\\
\section{Conclusions} 
We studied the one-dimensional Su-Schrieffer-Heeger(SSH) topological insulator with first and second neighbor-hoppings. In the simple SSH model, the TPT $v=w$ have states with ME $K\rightarrow 2$ that satisfy hybrid Bell conditions. The topological region $v<w$ always presents ME states, and the trivial region becomes disentangled $K\rightarrow 1$. \\
The extended SSH model with second neighbor hopping interaction $z\neq 0$  generates more ME states. The states with ME are contained in the topological region $v<w+z$ according to the winding numbers $\zeta=\pm1$.\\

In general, the Schmidt number is a good metric of bipartite entanglement that characterizes TPT and regions with ME. 
The chiral Hamiltonian for the SSH model presents a robust relationship between TPT and states with ME. More importantly, the Schmidt number as a metric of entangled allows us to characterize the winding number as a topological invariant in the SSH model. There is a strong relationship between the Schmidt number and localization of the wavefunction because in the TPT both quantities have a local maximum.\\

A new paradigm towards understanding the behavior of these topological material's properties opens the possibility to explore the hybrid nature of entangled states, as well as their potential application in quantum information processing. 
We only considered the effects mediated by hopping in a tight-binding model but would be desirable to study entanglement and topological phases for systems with more variety of interactions, couplings, long-range interaction, and spatial dimensions.  

\bibliography{references.bib}

\begin{thebibliography}{46}%
\makeatletter
\providecommand \@ifxundefined [1]{%
 \@ifx{#1\undefined}
}%
\providecommand \@ifnum [1]{%
 \ifnum #1\expandafter \@firstoftwo
 \else \expandafter \@secondoftwo
 \fi
}%
\providecommand \@ifx [1]{%
 \ifx #1\expandafter \@firstoftwo
 \else \expandafter \@secondoftwo
 \fi
}%
\providecommand \natexlab [1]{#1}%
\providecommand \enquote  [1]{``#1''}%
\providecommand \bibnamefont  [1]{#1}%
\providecommand \bibfnamefont [1]{#1}%
\providecommand \citenamefont [1]{#1}%
\providecommand \href@noop [0]{\@secondoftwo}%
\providecommand \href [0]{\begingroup \@sanitize@url \@href}%
\providecommand \@href[1]{\@@startlink{#1}\@@href}%
\providecommand \@@href[1]{\endgroup#1\@@endlink}%
\providecommand \@sanitize@url [0]{\catcode `\\12\catcode `\$12\catcode
  `\&12\catcode `\#12\catcode `\^12\catcode `\_12\catcode `\%12\relax}%
\providecommand \@@startlink[1]{}%
\providecommand \@@endlink[0]{}%
\providecommand \url  [0]{\begingroup\@sanitize@url \@url }%
\providecommand \@url [1]{\endgroup\@href {#1}{\urlprefix }}%
\providecommand \urlprefix  [0]{URL }%
\providecommand \Eprint [0]{\href }%
\providecommand \doibase [0]{https://doi.org/}%
\providecommand \selectlanguage [0]{\@gobble}%
\providecommand \bibinfo  [0]{\@secondoftwo}%
\providecommand \bibfield  [0]{\@secondoftwo}%
\providecommand \translation [1]{[#1]}%
\providecommand \BibitemOpen [0]{}%
\providecommand \bibitemStop [0]{}%
\providecommand \bibitemNoStop [0]{.\EOS\space}%
\providecommand \EOS [0]{\spacefactor3000\relax}%
\providecommand \BibitemShut  [1]{\csname bibitem#1\endcsname}%
\let\auto@bib@innerbib\@empty
\bibitem [{\citenamefont {Hasan}\ and\ \citenamefont {Kane}(2010)}]{Hasan2010}%
  \BibitemOpen
  \bibfield  {author} {\bibinfo {author} {\bibfnamefont {M.~Z.}\ \bibnamefont
  {Hasan}}\ and\ \bibinfo {author} {\bibfnamefont {C.~L.}\ \bibnamefont
  {Kane}},\ }\bibfield  {title} {\bibinfo {title} {Colloquium: Topological
  insulators},\ }\href {https://doi.org/10.1103/RevModPhys.82.3045} {\bibfield
  {journal} {\bibinfo  {journal} {Rev. Mod. Phys.}\ }\textbf {\bibinfo {volume}
  {82}},\ \bibinfo {pages} {3045} (\bibinfo {year} {2010})}\BibitemShut
  {NoStop}%
\bibitem [{\citenamefont {Qi}\ and\ \citenamefont {Zhang}(2011)}]{Xiao2011}%
  \BibitemOpen
  \bibfield  {author} {\bibinfo {author} {\bibfnamefont {X.-L.}\ \bibnamefont
  {Qi}}\ and\ \bibinfo {author} {\bibfnamefont {S.-C.}\ \bibnamefont {Zhang}},\
  }\bibfield  {title} {\bibinfo {title} {Topological insulators and
  superconductors},\ }\href {https://doi.org/10.1103/RevModPhys.83.1057}
  {\bibfield  {journal} {\bibinfo  {journal} {Rev. Mod. Phys.}\ }\textbf
  {\bibinfo {volume} {83}},\ \bibinfo {pages} {1057} (\bibinfo {year}
  {2011})}\BibitemShut {NoStop}%
\bibitem [{\citenamefont {Yu-Hang}\ \emph {et~al.}(2021)\citenamefont
  {Yu-Hang}, \citenamefont {Li}, \citenamefont {Ran},\ and\ \citenamefont
  {Cheng}}]{HANG2021}%
  \BibitemOpen
  \bibfield  {author} {\bibinfo {author} {\bibnamefont {Yu-Hang}}, \bibinfo
  {author} {\bibnamefont {Li}}, \bibinfo {author} {\bibnamefont {Ran}},\ and\
  \bibinfo {author} {\bibnamefont {Cheng}},\ }\bibfield  {title} {\bibinfo
  {title} {Magnonic su-schrieffer-heeger model in honeycomb ferromagnets},\
  }\href {https://doi.org/https://doi.org/10.1103/PhysRevB.103.014407}
  {\bibfield  {journal} {\bibinfo  {journal} {Phys. Rev. B}\ }\textbf {\bibinfo
  {volume} {103}},\ \bibinfo {pages} {014407} (\bibinfo {year}
  {2021})}\BibitemShut {NoStop}%
\bibitem [{\citenamefont {Böhling}\ \emph {et~al.}(2018)\citenamefont
  {Böhling}, \citenamefont {Engelhardt}, \citenamefont {Platero},\ and\
  \citenamefont {Schaller}}]{Platero2018}%
  \BibitemOpen
  \bibfield  {author} {\bibinfo {author} {\bibfnamefont {S.}~\bibnamefont
  {Böhling}}, \bibinfo {author} {\bibfnamefont {G.}~\bibnamefont
  {Engelhardt}}, \bibinfo {author} {\bibfnamefont {G.}~\bibnamefont
  {Platero}},\ and\ \bibinfo {author} {\bibfnamefont {G.}~\bibnamefont
  {Schaller}},\ }\bibfield  {title} {\bibinfo {title} {Thermoelectric
  performance of topological boundary modes},\ }\href
  {https://doi.org/https://doi.org/10.1103/PhysRevB.98.035132} {\bibfield
  {journal} {\bibinfo  {journal} {Phys. Rev. B}\ }\textbf {\bibinfo {volume}
  {98}},\ \bibinfo {pages} {035132} (\bibinfo {year} {2018})}\BibitemShut
  {NoStop}%
\bibitem [{\citenamefont {L.}\ \emph {et~al.}(2013)\citenamefont {L.},
  \citenamefont {Kane}, \citenamefont {C.},\ and\ \citenamefont
  {Lubensky}}]{Kane2014}%
  \BibitemOpen
  \bibfield  {author} {\bibinfo {author} {\bibfnamefont {C.}~\bibnamefont
  {L.}}, \bibinfo {author} {\bibnamefont {Kane}}, \bibinfo {author}
  {\bibfnamefont {T.}~\bibnamefont {C.}},\ and\ \bibinfo {author} {\bibnamefont
  {Lubensky}},\ }\bibfield  {title} {\bibinfo {title} {Topological boundary
  modes in isostatic lattices},\ }\href
  {https://doi.org/https://doi.org/10.1038/NPHYS2835} {\bibfield  {journal}
  {\bibinfo  {journal} {Nat.Phys.}\ }\textbf {\bibinfo {volume} {10}},\
  \bibinfo {pages} {39} (\bibinfo {year} {2013})}\BibitemShut {NoStop}%
\bibitem [{\citenamefont {Naumis}\ and\ \citenamefont
  {Roman-Taboada}(2014)}]{Naumis2014}%
  \BibitemOpen
  \bibfield  {author} {\bibinfo {author} {\bibfnamefont {G.~G.}\ \bibnamefont
  {Naumis}}\ and\ \bibinfo {author} {\bibfnamefont {P.}~\bibnamefont
  {Roman-Taboada}},\ }\bibfield  {title} {\bibinfo {title} {Mapping of strained
  graphene into one-dimensional hamiltonians: Quasicrystals and modulated
  crystals},\ }\href {https://doi.org/10.1103/PhysRevB.89.241404} {\bibfield
  {journal} {\bibinfo  {journal} {Phys. Rev. B}\ }\textbf {\bibinfo {volume}
  {89}},\ \bibinfo {pages} {241404} (\bibinfo {year} {2014})}\BibitemShut
  {NoStop}%
\bibitem [{\citenamefont {Manuel}\ \emph {et~al.}(2016)\citenamefont {Manuel},
  \citenamefont {Weber}, \citenamefont {F.}, \citenamefont {Assaad},\ and\
  \citenamefont {Hohenadler}}]{WEBER2016}%
  \BibitemOpen
  \bibfield  {author} {\bibinfo {author} {\bibnamefont {Manuel}}, \bibinfo
  {author} {\bibnamefont {Weber}}, \bibinfo {author} {\bibfnamefont
  {F.}~\bibnamefont {F.}}, \bibinfo {author} {\bibnamefont {Assaad}},\ and\
  \bibinfo {author} {\bibfnamefont {M.}~\bibnamefont {Hohenadler}},\ }\bibfield
   {title} {\bibinfo {title} {Thermodynamic and spectral properties of
  adiabatic peierls chains},\ }\href
  {https://doi.org/https://doi.org/10.1103/PhysRevB.94.155150} {\bibfield
  {journal} {\bibinfo  {journal} {Phys. Rev. B}\ }\textbf {\bibinfo {volume}
  {94}},\ \bibinfo {pages} {155150} (\bibinfo {year} {2016})}\BibitemShut
  {NoStop}%
\bibitem [{\citenamefont {Ziwei}\ \emph {et~al.}(2020)\citenamefont {Ziwei},
  \citenamefont {Fu}, \citenamefont {Nianzu}, \citenamefont {Fu}, \citenamefont
  {Huaiyuan}, \citenamefont {Zhang}, \citenamefont {Wang}, \citenamefont
  {Dong}, \citenamefont {Zhao}, \citenamefont {Shaolin},\ and\ \citenamefont
  {Ke}}]{SHAOLIN2020}%
  \BibitemOpen
  \bibfield  {author} {\bibinfo {author} {\bibnamefont {Ziwei}}, \bibinfo
  {author} {\bibnamefont {Fu}}, \bibinfo {author} {\bibnamefont {Nianzu}},
  \bibinfo {author} {\bibnamefont {Fu}}, \bibinfo {author} {\bibnamefont
  {Huaiyuan}}, \bibinfo {author} {\bibnamefont {Zhang}}, \bibinfo {author}
  {\bibfnamefont {Z.}~\bibnamefont {Wang}}, \bibinfo {author} {\bibnamefont
  {Dong}}, \bibinfo {author} {\bibnamefont {Zhao}}, \bibinfo {author}
  {\bibnamefont {Shaolin}},\ and\ \bibinfo {author} {\bibnamefont {Ke}},\
  }\bibfield  {title} {\bibinfo {title} {Extended ssh model in non-hermitian
  waveguides with alternating real and imaginary couplings},\ }\href
  {https://doi.org/https://doi.org/10.3390/app10103425} {\bibfield  {journal}
  {\bibinfo  {journal} {MDPI: Applied Sciences}\ }\textbf {\bibinfo {volume}
  {10}},\ \bibinfo {pages} {3425} (\bibinfo {year} {2020})}\BibitemShut
  {NoStop}%
\bibitem [{\citenamefont {A.}\ \emph {et~al.}(2010)\citenamefont {A.},
  \citenamefont {Neto}, \citenamefont {F.}, \citenamefont {Guinea},
  \citenamefont {N.}, \citenamefont {R.}, \citenamefont {Peres}, \citenamefont
  {K.}, \citenamefont {Novoselov},\ and\ \citenamefont {Geim}}]{Guinea2009}%
  \BibitemOpen
  \bibfield  {author} {\bibinfo {author} {\bibnamefont {A.}}, \bibinfo {author}
  {\bibnamefont {Neto}}, \bibinfo {author} {\bibnamefont {F.}}, \bibinfo
  {author} {\bibnamefont {Guinea}}, \bibinfo {author} {\bibnamefont {N.}},
  \bibinfo {author} {\bibfnamefont {M.}~\bibnamefont {R.}}, \bibinfo {author}
  {\bibnamefont {Peres}}, \bibinfo {author} {\bibnamefont {K.}}, \bibinfo
  {author} {\bibnamefont {Novoselov}},\ and\ \bibinfo {author} {\bibfnamefont
  {A.~K.}\ \bibnamefont {Geim}},\ }\bibfield  {title} {\bibinfo {title} {The
  electronic properties of graphene},\ }\href
  {https://doi.org/10.1103/RevModPhys.81.109} {\bibfield  {journal} {\bibinfo
  {journal} {Rev. Mod. Phys.}\ }\textbf {\bibinfo {volume} {81}},\ \bibinfo
  {pages} {109} (\bibinfo {year} {2010})}\BibitemShut {NoStop}%
\bibitem [{\citenamefont {Xiao-Long}\ \emph {et~al.}(2019)\citenamefont
  {Xiao-Long}, \citenamefont {Lü}, \citenamefont {Hang},\ and\ \citenamefont
  {Xie}}]{LONG2020}%
  \BibitemOpen
  \bibfield  {author} {\bibinfo {author} {\bibnamefont {Xiao-Long}}, \bibinfo
  {author} {\bibnamefont {Lü}}, \bibinfo {author} {\bibnamefont {Hang}},\ and\
  \bibinfo {author} {\bibnamefont {Xie}},\ }\bibfield  {title} {\bibinfo
  {title} {Topological phases and pumps in the su–schrieffer–heeger model
  periodically modulated in time},\ }\href
  {https://doi.org/https://doi.org/10.1088/1361-648X/ab3d72} {\bibfield
  {journal} {\bibinfo  {journal} {J. Phys.: Condens. Matter}\ }\textbf
  {\bibinfo {volume} {31}},\ \bibinfo {pages} {495401} (\bibinfo {year}
  {2019})}\BibitemShut {NoStop}%
\bibitem [{\citenamefont {Roman-Taboada}\ and\ \citenamefont
  {Naumis}(2015)}]{Taboada2015}%
  \BibitemOpen
  \bibfield  {author} {\bibinfo {author} {\bibfnamefont {P.}~\bibnamefont
  {Roman-Taboada}}\ and\ \bibinfo {author} {\bibfnamefont {G.~G.}\ \bibnamefont
  {Naumis}},\ }\bibfield  {title} {\bibinfo {title} {Spectral butterfly and
  electronic localization in rippled-graphene nanoribbons: Mapping onto
  effective one-dimensional chains},\ }\href
  {https://doi.org/10.1103/PhysRevB.92.035406} {\bibfield  {journal} {\bibinfo
  {journal} {Phys. Rev. B}\ }\textbf {\bibinfo {volume} {92}},\ \bibinfo
  {pages} {035406} (\bibinfo {year} {2015})}\BibitemShut {NoStop}%
\bibitem [{\citenamefont {Jay}\ \emph {et~al.}(2010)\citenamefont {Jay},
  \citenamefont {D.}, \citenamefont {Sau}, \citenamefont {Roman}, \citenamefont
  {Lutchyn}, \citenamefont {Tewari}, \citenamefont {Das},\ and\ \citenamefont
  {Sarma}}]{JAY2010}%
  \BibitemOpen
  \bibfield  {author} {\bibinfo {author} {\bibnamefont {Jay}}, \bibinfo
  {author} {\bibnamefont {D.}}, \bibinfo {author} {\bibnamefont {Sau}},
  \bibinfo {author} {\bibnamefont {Roman}}, \bibinfo {author} {\bibfnamefont
  {M.}~\bibnamefont {Lutchyn}}, \bibinfo {author} {\bibfnamefont
  {S.}~\bibnamefont {Tewari}}, \bibinfo {author} {\bibfnamefont
  {S.}~\bibnamefont {Das}},\ and\ \bibinfo {author} {\bibnamefont {Sarma}},\
  }\bibfield  {title} {\bibinfo {title} {Generic new platform for topological
  quantum computation using semiconductor heterostructures},\ }\href
  {https://doi.org/https://doi.org/10.1103/PhysRevLett.104.040502} {\bibfield
  {journal} {\bibinfo  {journal} {Phys. Rev. Lett.}\ }\textbf {\bibinfo
  {volume} {104}},\ \bibinfo {pages} {040502} (\bibinfo {year}
  {2010})}\BibitemShut {NoStop}%
\bibitem [{\citenamefont {Naumis}\ \emph {et~al.}(2021)\citenamefont {Naumis},
  \citenamefont {Navarro-Labastida}, \citenamefont {Aguilar-M\'endez},\ and\
  \citenamefont {Espinosa-Champo}}]{Navarro2021}%
  \BibitemOpen
  \bibfield  {author} {\bibinfo {author} {\bibfnamefont {G.~G.}\ \bibnamefont
  {Naumis}}, \bibinfo {author} {\bibfnamefont {L.~A.}\ \bibnamefont
  {Navarro-Labastida}}, \bibinfo {author} {\bibfnamefont {E.}~\bibnamefont
  {Aguilar-M\'endez}},\ and\ \bibinfo {author} {\bibfnamefont {A.}~\bibnamefont
  {Espinosa-Champo}},\ }\bibfield  {title} {\bibinfo {title} {Reduction of the
  twisted bilayer graphene chiral hamiltonian into a
  $2\ifmmode\times\else\texttimes\fi{}2$ matrix operator and physical origin of
  flat bands at magic angles},\ }\href
  {https://doi.org/10.1103/PhysRevB.103.245418} {\bibfield  {journal} {\bibinfo
   {journal} {Phys. Rev. B}\ }\textbf {\bibinfo {volume} {103}},\ \bibinfo
  {pages} {245418} (\bibinfo {year} {2021})}\BibitemShut {NoStop}%
\bibitem [{\citenamefont {Chen}\ and\ \citenamefont {Li}(2010)}]{YIXIN2010}%
  \BibitemOpen
  \bibfield  {author} {\bibinfo {author} {\bibfnamefont {Y.-X.}\ \bibnamefont
  {Chen}}\ and\ \bibinfo {author} {\bibfnamefont {S.-W.}\ \bibnamefont {Li}},\
  }\bibfield  {title} {\bibinfo {title} {Quantum correlations in topological
  quantum phase transitions},\ }\href
  {https://doi.org/https://doi.org/10.1103/PhysRevA.81.032120} {\bibfield
  {journal} {\bibinfo  {journal} {Phys. Rev. A}\ }\textbf {\bibinfo {volume}
  {81}},\ \bibinfo {pages} {032120} (\bibinfo {year} {2010})}\BibitemShut
  {NoStop}%
\bibitem [{\citenamefont {Hauke}\ \emph {et~al.}(2016)\citenamefont {Hauke},
  \citenamefont {Heyl}, \citenamefont {Tagliacozzo},\ and\ \citenamefont
  {Zoller}}]{Hauke2014}%
  \BibitemOpen
  \bibfield  {author} {\bibinfo {author} {\bibfnamefont {P.}~\bibnamefont
  {Hauke}}, \bibinfo {author} {\bibfnamefont {M.}~\bibnamefont {Heyl}},
  \bibinfo {author} {\bibfnamefont {L.}~\bibnamefont {Tagliacozzo}},\ and\
  \bibinfo {author} {\bibfnamefont {P.}~\bibnamefont {Zoller}},\ }\bibfield
  {title} {\bibinfo {title} {Measuring multipartite entanglement through
  dynamic susceptibilities},\ }\href
  {https://doi.org/https://doi.org/10.1038/NPHYS3700} {\bibfield  {journal}
  {\bibinfo  {journal} {Nature Physics}\ }\textbf {\bibinfo {volume} {12}},\
  \bibinfo {pages} {782} (\bibinfo {year} {2016})}\BibitemShut {NoStop}%
\bibitem [{\citenamefont {Xiao}\ \emph {et~al.}(2010)\citenamefont {Xiao},
  \citenamefont {Chang},\ and\ \citenamefont {Niu}}]{NIU2010}%
  \BibitemOpen
  \bibfield  {author} {\bibinfo {author} {\bibfnamefont {D.}~\bibnamefont
  {Xiao}}, \bibinfo {author} {\bibfnamefont {M.-C.}\ \bibnamefont {Chang}},\
  and\ \bibinfo {author} {\bibfnamefont {Q.}~\bibnamefont {Niu}},\ }\bibfield
  {title} {\bibinfo {title} {Berry phase effects on electronic properties},\
  }\href {https://doi.org/https://doi.org/10.1103/RevModPhys.82.1959}
  {\bibfield  {journal} {\bibinfo  {journal} {Rev. Mod. Phys.}\ }\textbf
  {\bibinfo {volume} {82}},\ \bibinfo {pages} {1959} (\bibinfo {year}
  {2010})}\BibitemShut {NoStop}%
\bibitem [{\citenamefont {Li}\ \emph {et~al.}(2014)\citenamefont {Li},
  \citenamefont {Yu}, \citenamefont {Lin},\ and\ \citenamefont
  {You}}]{Jun2014}%
  \BibitemOpen
  \bibfield  {author} {\bibinfo {author} {\bibfnamefont {J.}~\bibnamefont
  {Li}}, \bibinfo {author} {\bibfnamefont {T.}~\bibnamefont {Yu}}, \bibinfo
  {author} {\bibfnamefont {H.-Q.}\ \bibnamefont {Lin}},\ and\ \bibinfo {author}
  {\bibfnamefont {J.~Q.}\ \bibnamefont {You}},\ }\bibfield  {title} {\bibinfo
  {title} {Probing the non-locality of majorana fermions via quantum
  correlations},\ }\href {https://doi.org/https://doi.org/10.1038/srep04930}
  {\bibfield  {journal} {\bibinfo  {journal} {Nature Scientific Reports}\
  }\textbf {\bibinfo {volume} {408}},\ \bibinfo {pages} {4930} (\bibinfo {year}
  {2014})}\BibitemShut {NoStop}%
\bibitem [{\citenamefont {Yu}\ and\ \citenamefont {Eberly}(2006)}]{YU2006}%
  \BibitemOpen
  \bibfield  {author} {\bibinfo {author} {\bibfnamefont {T.}~\bibnamefont
  {Yu}}\ and\ \bibinfo {author} {\bibfnamefont {J.~H.}\ \bibnamefont
  {Eberly}},\ }\bibfield  {title} {\bibinfo {title} {Quantum open system
  theory: Bipartite aspects},\ }\href
  {https://doi.org/https://doi.org/10.1103/PhysRevLett.97.140403} {\bibfield
  {journal} {\bibinfo  {journal} {Phys. Rev. Lett.}\ }\textbf {\bibinfo
  {volume} {97}},\ \bibinfo {pages} {140403} (\bibinfo {year}
  {2006})}\BibitemShut {NoStop}%
\bibitem [{\citenamefont {Zeng}\ \emph {et~al.}(2019)\citenamefont {Zeng},
  \citenamefont {Shi}, \citenamefont {Zhou}, \citenamefont {Wang},
  \citenamefont {Liu},\ and\ \citenamefont {Hu}}]{Ping2019}%
  \BibitemOpen
  \bibfield  {author} {\bibinfo {author} {\bibfnamefont {S.-P.}\ \bibnamefont
  {Zeng}}, \bibinfo {author} {\bibfnamefont {H.-L.}\ \bibnamefont {Shi}},
  \bibinfo {author} {\bibfnamefont {X.}~\bibnamefont {Zhou}}, \bibinfo {author}
  {\bibfnamefont {X.-H.}\ \bibnamefont {Wang}}, \bibinfo {author}
  {\bibfnamefont {S.-Y.}\ \bibnamefont {Liu}},\ and\ \bibinfo {author}
  {\bibfnamefont {M.-L.}\ \bibnamefont {Hu}},\ }\bibfield  {title} {\bibinfo
  {title} {Protecting quantum correlations of the xxz model by topological
  boundary conditions},\ }\href
  {https://doi.org/https://doi.org/10.1038/s41598-018-37474-x} {\bibfield
  {journal} {\bibinfo  {journal} {Nature Scientific Reports}\ }\textbf
  {\bibinfo {volume} {1083}},\ \bibinfo {pages} {2389} (\bibinfo {year}
  {2019})}\BibitemShut {NoStop}%
\bibitem [{\citenamefont {Cho}\ and\ \citenamefont {Kim}(2017)}]{Jaeyoon2017}%
  \BibitemOpen
  \bibfield  {author} {\bibinfo {author} {\bibfnamefont {J.}~\bibnamefont
  {Cho}}\ and\ \bibinfo {author} {\bibfnamefont {K.~W.}\ \bibnamefont {Kim}},\
  }\bibfield  {title} {\bibinfo {title} {Quantum phase transition and
  entanglement in topological quantum wires},\ }\href
  {https://doi.org/https://doi.org/10.1038/s41598-017-02717-w} {\bibfield
  {journal} {\bibinfo  {journal} {Nature Scientific Reports}\ }\textbf
  {\bibinfo {volume} {7}},\ \bibinfo {pages} {2745} (\bibinfo {year}
  {2017})}\BibitemShut {NoStop}%
\bibitem [{\citenamefont {{Morita}}\ \emph {et~al.}(2021)\citenamefont
  {{Morita}}, \citenamefont {{Sota}},\ and\ \citenamefont
  {{Tohyama}}}]{2021TakamiT}%
  \BibitemOpen
  \bibfield  {author} {\bibinfo {author} {\bibfnamefont {K.}~\bibnamefont
  {{Morita}}}, \bibinfo {author} {\bibfnamefont {S.}~\bibnamefont {{Sota}}},\
  and\ \bibinfo {author} {\bibfnamefont {T.}~\bibnamefont {{Tohyama}}},\
  }\bibfield  {title} {\bibinfo {title} {{Magnetic phase diagrams of the
  spin-$\frac{1}{2}$ Heisenberg model on a kagome-strip chain: Emergence of a
  Haldane phase}},\ }\href@noop {} {\bibfield  {journal} {\bibinfo  {journal}
  {arXiv e-prints}\ ,\ \bibinfo {eid} {arXiv:2108.12584}} (\bibinfo {year}
  {2021})},\ \Eprint {https://arxiv.org/abs/2108.12584} {arXiv:2108.12584
  [cond-mat.str-el]} \BibitemShut {NoStop}%
\bibitem [{\citenamefont {Brody}\ and\ \citenamefont
  {Hughston}(2001)}]{Brody2001}%
  \BibitemOpen
  \bibfield  {author} {\bibinfo {author} {\bibfnamefont {D.~C.}\ \bibnamefont
  {Brody}}\ and\ \bibinfo {author} {\bibfnamefont {L.~P.}\ \bibnamefont
  {Hughston}},\ }\bibfield  {title} {\bibinfo {title} {Geometric quantum
  mechanics},\ }\href {https://doi.org/10.1016/S0393-0440(00)00052-8}
  {\bibfield  {journal} {\bibinfo  {journal} {Journal of Geometry and Physics}\
  }\textbf {\bibinfo {volume} {38}},\ \bibinfo {pages} {19} (\bibinfo {year}
  {2001})}\BibitemShut {NoStop}%
\bibitem [{\citenamefont {Horodecki}\ \emph {et~al.}(2009)\citenamefont
  {Horodecki}, \citenamefont {Horodecki}, \citenamefont {Horodecki},\ and\
  \citenamefont {Horodecki}}]{Horodecki2009}%
  \BibitemOpen
  \bibfield  {author} {\bibinfo {author} {\bibfnamefont {R.}~\bibnamefont
  {Horodecki}}, \bibinfo {author} {\bibfnamefont {P.}~\bibnamefont
  {Horodecki}}, \bibinfo {author} {\bibfnamefont {M.}~\bibnamefont
  {Horodecki}},\ and\ \bibinfo {author} {\bibfnamefont {K.}~\bibnamefont
  {Horodecki}},\ }\bibfield  {title} {\bibinfo {title} {Quantum entanglement},\
  }\href {https://doi.org/https://doi.org/10.1103/RevModPhys.81.865} {\bibfield
   {journal} {\bibinfo  {journal} {Rev. Mod. Phys.}\ }\textbf {\bibinfo
  {volume} {81}},\ \bibinfo {pages} {865} (\bibinfo {year} {2009})}\BibitemShut
  {NoStop}%
\bibitem [{\citenamefont {Valerio}\ \emph {et~al.}(2009)\citenamefont
  {Valerio}, \citenamefont {Scarani}, \citenamefont {Helle}, \citenamefont
  {Bechmann-Pasquinucci}, \citenamefont {J.}, \citenamefont {Cerf},
  \citenamefont {Miloslav}, \citenamefont {Dušek}, \citenamefont {Norbert},
  \citenamefont {Lütkenhaus},\ and\ \citenamefont {Peev}}]{VALERIO2009}%
  \BibitemOpen
  \bibfield  {author} {\bibinfo {author} {\bibnamefont {Valerio}}, \bibinfo
  {author} {\bibnamefont {Scarani}}, \bibinfo {author} {\bibnamefont {Helle}},
  \bibinfo {author} {\bibnamefont {Bechmann-Pasquinucci}}, \bibinfo {author}
  {\bibfnamefont {N.}~\bibnamefont {J.}}, \bibinfo {author} {\bibnamefont
  {Cerf}}, \bibinfo {author} {\bibnamefont {Miloslav}}, \bibinfo {author}
  {\bibnamefont {Dušek}}, \bibinfo {author} {\bibnamefont {Norbert}}, \bibinfo
  {author} {\bibfnamefont {M.}~\bibnamefont {Lütkenhaus}},\ and\ \bibinfo
  {author} {\bibnamefont {Peev}},\ }\bibfield  {title} {\bibinfo {title} {The
  security of practical quantum key distribution},\ }\href
  {https://doi.org/https://doi.org/10.1103/RevModPhys.81.1301} {\bibfield
  {journal} {\bibinfo  {journal} {Rev. Mod. Phys.}\ }\textbf {\bibinfo {volume}
  {81}},\ \bibinfo {pages} {1301} (\bibinfo {year} {2009})}\BibitemShut
  {NoStop}%
\bibitem [{\citenamefont {Pan}\ \emph {et~al.}(2012)\citenamefont {Pan},
  \citenamefont {Chen}, \citenamefont {Lu}, \citenamefont {Weinfurter},
  \citenamefont {Zeilinger},\ and\ \citenamefont {Żukowski}}]{JIAN2012}%
  \BibitemOpen
  \bibfield  {author} {\bibinfo {author} {\bibfnamefont {J.-W.}\ \bibnamefont
  {Pan}}, \bibinfo {author} {\bibfnamefont {Z.-B.}\ \bibnamefont {Chen}},
  \bibinfo {author} {\bibfnamefont {C.-Y.}\ \bibnamefont {Lu}}, \bibinfo
  {author} {\bibfnamefont {H.}~\bibnamefont {Weinfurter}}, \bibinfo {author}
  {\bibfnamefont {A.}~\bibnamefont {Zeilinger}},\ and\ \bibinfo {author}
  {\bibfnamefont {M.}~\bibnamefont {Żukowski}},\ }\bibfield  {title} {\bibinfo
  {title} {Multiphoton entanglement and interferometry},\ }\href
  {https://doi.org/https://doi.org/10.1103/RevModPhys.84.777} {\bibfield
  {journal} {\bibinfo  {journal} {Rev. Mod. Phys.}\ }\textbf {\bibinfo {volume}
  {84}},\ \bibinfo {pages} {177} (\bibinfo {year} {2012})}\BibitemShut
  {NoStop}%
\bibitem [{\citenamefont {Nicolas}\ \emph {et~al.}(2012)\citenamefont
  {Nicolas}, \citenamefont {Gisin}, \citenamefont {Grégoire}, \citenamefont
  {Ribordy}, \citenamefont {Tittel}, \citenamefont {Hugo},\ and\ \citenamefont
  {Zbinden}}]{NICOLAS2012}%
  \BibitemOpen
  \bibfield  {author} {\bibinfo {author} {\bibnamefont {Nicolas}}, \bibinfo
  {author} {\bibnamefont {Gisin}}, \bibinfo {author} {\bibnamefont
  {Grégoire}}, \bibinfo {author} {\bibnamefont {Ribordy}}, \bibinfo {author}
  {\bibfnamefont {W.}~\bibnamefont {Tittel}}, \bibinfo {author} {\bibnamefont
  {Hugo}},\ and\ \bibinfo {author} {\bibnamefont {Zbinden}},\ }\bibfield
  {title} {\bibinfo {title} {Quantum cryptography},\ }\href
  {https://doi.org/https://doi.org/10.1103/RevModPhys.74.145} {\bibfield
  {journal} {\bibinfo  {journal} {Rev. Mod. Phys.}\ }\textbf {\bibinfo {volume}
  {74}},\ \bibinfo {pages} {145} (\bibinfo {year} {2012})}\BibitemShut
  {NoStop}%
\bibitem [{\citenamefont {Ota}\ \emph {et~al.}(2020)\citenamefont {Ota},
  \citenamefont {Takata}, \citenamefont {Ozawa}, \citenamefont {Amo},
  \citenamefont {Jia}, \citenamefont {Kante}, \citenamefont {Notomi},
  \citenamefont {Arakawa},\ and\ \citenamefont {Iwamoto}}]{Ozawa2020}%
  \BibitemOpen
  \bibfield  {author} {\bibinfo {author} {\bibfnamefont {Y.}~\bibnamefont
  {Ota}}, \bibinfo {author} {\bibfnamefont {K.}~\bibnamefont {Takata}},
  \bibinfo {author} {\bibfnamefont {T.}~\bibnamefont {Ozawa}}, \bibinfo
  {author} {\bibfnamefont {A.}~\bibnamefont {Amo}}, \bibinfo {author}
  {\bibfnamefont {Z.}~\bibnamefont {Jia}}, \bibinfo {author} {\bibfnamefont
  {B.}~\bibnamefont {Kante}}, \bibinfo {author} {\bibfnamefont
  {M.}~\bibnamefont {Notomi}}, \bibinfo {author} {\bibfnamefont
  {Y.}~\bibnamefont {Arakawa}},\ and\ \bibinfo {author} {\bibfnamefont
  {S.}~\bibnamefont {Iwamoto}},\ }\bibfield  {title} {\bibinfo {title} {Active
  topological photonics},\ }\href
  {https://doi.org/doi:10.1515/nanoph-2019-0376} {\bibfield  {journal}
  {\bibinfo  {journal} {Nanophotonics}\ }\textbf {\bibinfo {volume} {9}},\
  \bibinfo {pages} {547} (\bibinfo {year} {2020})}\BibitemShut {NoStop}%
\bibitem [{\citenamefont {Yang}\ \emph {et~al.}(2019)\citenamefont {Yang},
  \citenamefont {Y}, \citenamefont {Gao}, \citenamefont {Z}, \citenamefont
  {Xue}, \citenamefont {H},\ and\ \citenamefont {et~al.}}]{YANG2019}%
  \BibitemOpen
  \bibfield  {author} {\bibinfo {author} {\bibnamefont {Yang}}, \bibinfo
  {author} {\bibnamefont {Y}}, \bibinfo {author} {\bibnamefont {Gao}}, \bibinfo
  {author} {\bibnamefont {Z}}, \bibinfo {author} {\bibnamefont {Xue}}, \bibinfo
  {author} {\bibnamefont {H}},\ and\ \bibinfo {author} {\bibnamefont
  {et~al.}},\ }\bibfield  {title} {\bibinfo {title} {Realization of a
  three-dimensional photonic topological insulator},\ }\href
  {https://doi.org/10.1038/s41586-018-0829-0} {\bibfield  {journal} {\bibinfo
  {journal} {Nature}\ }\textbf {\bibinfo {volume} {565}},\ \bibinfo {pages}
  {622} (\bibinfo {year} {2019})}\BibitemShut {NoStop}%
\bibitem [{\citenamefont {de~Oliveira}\ \emph {et~al.}(2021)\citenamefont
  {de~Oliveira}, \citenamefont {Santos}, \citenamefont {Jesus-Silva},\ and\
  \citenamefont {Fonseca}}]{2021Oliveira}%
  \BibitemOpen
  \bibfield  {author} {\bibinfo {author} {\bibfnamefont {J.~M.}\ \bibnamefont
  {de~Oliveira}}, \bibinfo {author} {\bibfnamefont {L.~M.~S.}\ \bibnamefont
  {Santos}}, \bibinfo {author} {\bibfnamefont {A.~J.}\ \bibnamefont
  {Jesus-Silva}},\ and\ \bibinfo {author} {\bibfnamefont {E.~J.~S.}\
  \bibnamefont {Fonseca}},\ }\bibfield  {title} {\bibinfo {title} {Tunable
  generation and propagation of vortex beams in a photonic chip},\ }\href
  {https://doi.org/10.1103/PhysRevA.104.L061501} {\bibfield  {journal}
  {\bibinfo  {journal} {Phys. Rev. A}\ }\textbf {\bibinfo {volume} {104}},\
  \bibinfo {pages} {L061501} (\bibinfo {year} {2021})}\BibitemShut {NoStop}%
\bibitem [{\citenamefont {Yang}\ \emph {et~al.}(2020)\citenamefont {Yang},
  \citenamefont {Gao}, \citenamefont {Feng}, \citenamefont {Huang},
  \citenamefont {Zhou}, \citenamefont {Yang}, \citenamefont {Chong},\ and\
  \citenamefont {Zhang}}]{2020YANGG}%
  \BibitemOpen
  \bibfield  {author} {\bibinfo {author} {\bibfnamefont {Y.}~\bibnamefont
  {Yang}}, \bibinfo {author} {\bibfnamefont {Z.}~\bibnamefont {Gao}}, \bibinfo
  {author} {\bibfnamefont {X.}~\bibnamefont {Feng}}, \bibinfo {author}
  {\bibfnamefont {Y.-X.}\ \bibnamefont {Huang}}, \bibinfo {author}
  {\bibfnamefont {P.}~\bibnamefont {Zhou}}, \bibinfo {author} {\bibfnamefont
  {S.~A.}\ \bibnamefont {Yang}}, \bibinfo {author} {\bibfnamefont
  {Y.}~\bibnamefont {Chong}},\ and\ \bibinfo {author} {\bibfnamefont
  {B.}~\bibnamefont {Zhang}},\ }\bibfield  {title} {\bibinfo {title} {Ideal
  unconventional weyl point in a chiral photonic metamaterial},\ }\href
  {https://doi.org/10.1103/PhysRevLett.125.143001} {\bibfield  {journal}
  {\bibinfo  {journal} {Phys. Rev. Lett.}\ }\textbf {\bibinfo {volume} {125}},\
  \bibinfo {pages} {143001} (\bibinfo {year} {2020})}\BibitemShut {NoStop}%
\bibitem [{\citenamefont {Yang}\ \emph {et~al.}(2018)\citenamefont {Yang},
  \citenamefont {Xu}, \citenamefont {Xu}, \citenamefont {Wang}, \citenamefont
  {Jiang}, \citenamefont {Hu},\ and\ \citenamefont {Hang}}]{2018Yuting}%
  \BibitemOpen
  \bibfield  {author} {\bibinfo {author} {\bibfnamefont {Y.}~\bibnamefont
  {Yang}}, \bibinfo {author} {\bibfnamefont {Y.~F.}\ \bibnamefont {Xu}},
  \bibinfo {author} {\bibfnamefont {T.}~\bibnamefont {Xu}}, \bibinfo {author}
  {\bibfnamefont {H.-X.}\ \bibnamefont {Wang}}, \bibinfo {author}
  {\bibfnamefont {J.-H.}\ \bibnamefont {Jiang}}, \bibinfo {author}
  {\bibfnamefont {X.}~\bibnamefont {Hu}},\ and\ \bibinfo {author}
  {\bibfnamefont {Z.~H.}\ \bibnamefont {Hang}},\ }\bibfield  {title} {\bibinfo
  {title} {Visualization of a unidirectional electromagnetic waveguide using
  topological photonic crystals made of dielectric materials},\ }\href
  {https://doi.org/10.1103/PhysRevLett.120.217401} {\bibfield  {journal}
  {\bibinfo  {journal} {Phys. Rev. Lett.}\ }\textbf {\bibinfo {volume} {120}},\
  \bibinfo {pages} {217401} (\bibinfo {year} {2018})}\BibitemShut {NoStop}%
\bibitem [{\citenamefont {Su}\ \emph {et~al.}(1980)\citenamefont {Su},
  \citenamefont {Schrieffer},\ and\ \citenamefont {Heeger}}]{Schrieffer1980}%
  \BibitemOpen
  \bibfield  {author} {\bibinfo {author} {\bibnamefont {Su}}, \bibinfo {author}
  {\bibnamefont {Schrieffer}},\ and\ \bibinfo {author} {\bibnamefont
  {Heeger}},\ }\bibfield  {title} {\bibinfo {title} {Soliton excitations in
  polyacetylene},\ }\href
  {https://doi.org/https://doi.org/10.1103/PhysRevB.22.2099} {\bibfield
  {journal} {\bibinfo  {journal} {Phys. Rev. B}\ }\textbf {\bibinfo {volume}
  {22}},\ \bibinfo {pages} {2099} (\bibinfo {year} {1980})}\BibitemShut
  {NoStop}%
\bibitem [{\citenamefont {Ekert}\ and\ \citenamefont
  {Knight}(1995)}]{Knight2006}%
  \BibitemOpen
  \bibfield  {author} {\bibinfo {author} {\bibfnamefont {A.}~\bibnamefont
  {Ekert}}\ and\ \bibinfo {author} {\bibfnamefont {P.~L.}\ \bibnamefont
  {Knight}},\ }\bibfield  {title} {\bibinfo {title} {Entangled quantum systems
  and the schmidt decomposition},\ }\href {https://doi.org/10.1119/1.17904}
  {\bibfield  {journal} {\bibinfo  {journal} {Am. J. Phys.}\ }\textbf {\bibinfo
  {volume} {63}},\ \bibinfo {pages} {415} (\bibinfo {year} {1995})}\BibitemShut
  {NoStop}%
\bibitem [{\citenamefont {Sperling}\ and\ \citenamefont
  {Vogel}(1995)}]{Sperling2011}%
  \BibitemOpen
  \bibfield  {author} {\bibinfo {author} {\bibfnamefont {J.}~\bibnamefont
  {Sperling}}\ and\ \bibinfo {author} {\bibfnamefont {W.}~\bibnamefont
  {Vogel}},\ }\bibfield  {title} {\bibinfo {title} {Entangled quantum systems
  and the schmidt decomposition},\ }\href
  {https://doi.org/https://doi.org/10.1103/PhysRevA.83.042315} {\bibfield
  {journal} {\bibinfo  {journal} {Phys. Rev. A}\ }\textbf {\bibinfo {volume}
  {83}},\ \bibinfo {pages} {042315} (\bibinfo {year} {1995})}\BibitemShut
  {NoStop}%
\bibitem [{\citenamefont {Daichi}\ \emph {et~al.}(2019)\citenamefont {Daichi},
  \citenamefont {Obana}, \citenamefont {Feng}, \citenamefont {Liu},
  \citenamefont {Katsunori},\ and\ \citenamefont {Wakabayashi}}]{Obana2019}%
  \BibitemOpen
  \bibfield  {author} {\bibinfo {author} {\bibnamefont {Daichi}}, \bibinfo
  {author} {\bibnamefont {Obana}}, \bibinfo {author} {\bibnamefont {Feng}},
  \bibinfo {author} {\bibnamefont {Liu}}, \bibinfo {author} {\bibnamefont
  {Katsunori}},\ and\ \bibinfo {author} {\bibnamefont {Wakabayashi}},\
  }\bibfield  {title} {\bibinfo {title} {Topological edge states in the
  su-schrieffer-heeger model},\ }\href
  {https://doi.org/10.1103/PhysRevB.100.075437} {\bibfield  {journal} {\bibinfo
   {journal} {Phys. Rev. B}\ }\textbf {\bibinfo {volume} {100}},\ \bibinfo
  {pages} {075437} (\bibinfo {year} {2019})}\BibitemShut {NoStop}%
\bibitem [{\citenamefont {Kuno}(2019)}]{Kuno2019}%
  \BibitemOpen
  \bibfield  {author} {\bibinfo {author} {\bibfnamefont {Y.}~\bibnamefont
  {Kuno}},\ }\bibfield  {title} {\bibinfo {title} {Phase structure of the
  interacting su-schrieffer-heeger model and the relationship with the
  gross-neveu model on lattice},\ }\href
  {https://doi.org/10.1103/PhysRevB.99.064105} {\bibfield  {journal} {\bibinfo
  {journal} {Phys. Rev. B}\ }\textbf {\bibinfo {volume} {99}},\ \bibinfo
  {pages} {064105} (\bibinfo {year} {2019})}\BibitemShut {NoStop}%
\bibitem [{\citenamefont {Li}\ and\ \citenamefont
  {Miroshnichenko}(2018)}]{CHAO2010}%
  \BibitemOpen
  \bibfield  {author} {\bibinfo {author} {\bibfnamefont {C.}~\bibnamefont
  {Li}}\ and\ \bibinfo {author} {\bibfnamefont {A.~E.}\ \bibnamefont
  {Miroshnichenko}},\ }\bibfield  {title} {\bibinfo {title} {Extended ssh
  model: Non-local couplings and non-monotonous edge states},\ }\href
  {https://doi.org/https://doi.org/10.3390/physics1010002} {\bibfield
  {journal} {\bibinfo  {journal} {MDPI}\ }\textbf {\bibinfo {volume} {1}},\
  \bibinfo {pages} {2} (\bibinfo {year} {2018})}\BibitemShut {NoStop}%
\bibitem [{\citenamefont {Resta}(2000)}]{RESTA2000}%
  \BibitemOpen
  \bibfield  {author} {\bibinfo {author} {\bibfnamefont {R.}~\bibnamefont
  {Resta}},\ }\bibfield  {title} {\bibinfo {title} {Manifestations of berry's
  phase in molecules and condensed matter},\ }\href
  {https://doi.org/https://doi.org/10.1088/0953-8984/12/9/201} {\bibfield
  {journal} {\bibinfo  {journal} {J. Phys. Condens. Matter}\ }\textbf {\bibinfo
  {volume} {12}},\ \bibinfo {pages} {9} (\bibinfo {year} {2000})}\BibitemShut
  {NoStop}%
\bibitem [{\citenamefont {Simon}(1983)}]{BARRY1983}%
  \BibitemOpen
  \bibfield  {author} {\bibinfo {author} {\bibfnamefont {B.}~\bibnamefont
  {Simon}},\ }\bibfield  {title} {\bibinfo {title} {Holonomy, the quantum
  adiabatic theorem, and berry's phase},\ }\href
  {https://doi.org/https://doi.org/10.1103/PhysRevLett.51.2167} {\bibfield
  {journal} {\bibinfo  {journal} {Phys. Rev. Lett.}\ }\textbf {\bibinfo
  {volume} {51}},\ \bibinfo {pages} {2167} (\bibinfo {year}
  {1983})}\BibitemShut {NoStop}%
\bibitem [{\citenamefont {Resta}(1998)}]{RESTA1998}%
  \BibitemOpen
  \bibfield  {author} {\bibinfo {author} {\bibfnamefont {R.}~\bibnamefont
  {Resta}},\ }\bibfield  {title} {\bibinfo {title} {Quantum-mechanical position
  operator in extended systems},\ }\href
  {https://doi.org/10.1103/PhysRevLett.80.1800} {\bibfield  {journal} {\bibinfo
   {journal} {Phys. Rev. Lett.}\ }\textbf {\bibinfo {volume} {80}},\ \bibinfo
  {pages} {1800} (\bibinfo {year} {1998})}\BibitemShut {NoStop}%
\bibitem [{\citenamefont {Ortiz}\ and\ \citenamefont
  {Martin}(1994)}]{MARTIN1994}%
  \BibitemOpen
  \bibfield  {author} {\bibinfo {author} {\bibfnamefont {G.}~\bibnamefont
  {Ortiz}}\ and\ \bibinfo {author} {\bibfnamefont {R.~M.}\ \bibnamefont
  {Martin}},\ }\bibfield  {title} {\bibinfo {title} {Macroscopic polarization
  as a geometric quantum phase: Many-body formulation},\ }\href
  {https://doi.org/https://doi.org/10.1103/PhysRevB.49.14202} {\bibfield
  {journal} {\bibinfo  {journal} {Phys. Rev. B}\ }\textbf {\bibinfo {volume}
  {49}},\ \bibinfo {pages} {14202} (\bibinfo {year} {1994})}\BibitemShut
  {NoStop}%
\bibitem [{\citenamefont {Li}\ and\ \citenamefont
  {Fleischhauer}(2017)}]{MICHAEL2017}%
  \BibitemOpen
  \bibfield  {author} {\bibinfo {author} {\bibfnamefont {R.}~\bibnamefont
  {Li}}\ and\ \bibinfo {author} {\bibfnamefont {M.}~\bibnamefont
  {Fleischhauer}},\ }\bibfield  {title} {\bibinfo {title} {Finite-size
  corrections to quantized particle transport in topological charge pumps},\
  }\href {https://doi.org/https://doi.org/10.1103/PhysRevB.96.085444}
  {\bibfield  {journal} {\bibinfo  {journal} {Phys. Rev. B}\ }\textbf {\bibinfo
  {volume} {96}},\ \bibinfo {pages} {085444} (\bibinfo {year}
  {2017})}\BibitemShut {NoStop}%
\bibitem [{\citenamefont {Hetényi}\ \emph {et~al.}(2021)\citenamefont
  {Hetényi}, \citenamefont {Pulcu},\ and\ \citenamefont
  {Doğan}}]{balazss2021}%
  \BibitemOpen
  \bibfield  {author} {\bibinfo {author} {\bibfnamefont {B.}~\bibnamefont
  {Hetényi}}, \bibinfo {author} {\bibfnamefont {Y.}~\bibnamefont {Pulcu}},\
  and\ \bibinfo {author} {\bibfnamefont {S.}~\bibnamefont {Doğan}},\
  }\bibfield  {title} {\bibinfo {title} {Calculating the polarization in
  bipartite lattice models: Application to an extended su-schrieffer-heeger
  model},\ }\href {https://doi.org/10.1103/PhysRevB.103.075117} {\bibfield
  {journal} {\bibinfo  {journal} {Phys. Rev. Lett.}\ }\textbf {\bibinfo
  {volume} {103}},\ \bibinfo {pages} {075117} (\bibinfo {year}
  {2021})}\BibitemShut {NoStop}%
\bibitem [{\citenamefont {Bogdanov}\ \emph {et~al.}(2007)\citenamefont
  {Bogdanov}, \citenamefont {A.Y.}, \citenamefont {Y.I.},\ and\ \citenamefont
  {Valiev}}]{Bagdanov2007}%
  \BibitemOpen
  \bibfield  {author} {\bibinfo {author} {\bibnamefont {Bogdanov}}, \bibinfo
  {author} {\bibnamefont {A.Y.}}, \bibinfo {author} {\bibnamefont {Y.I.}},\
  and\ \bibinfo {author} {\bibfnamefont {K.}~\bibnamefont {Valiev}},\
  }\bibfield  {title} {\bibinfo {title} {Schmidt information and entanglement
  of quantum systems},\ }\href
  {https://doi.org/https://doi.org/10.3103/S0278641907010074} {\bibfield
  {journal} {\bibinfo  {journal} {Moscow Univ. Comput. Math. Cybern.}\ }\textbf
  {\bibinfo {volume} {31}},\ \bibinfo {pages} {33} (\bibinfo {year}
  {2007})}\BibitemShut {NoStop}%
\bibitem [{\citenamefont {Eberly}(2006)}]{Eberley2006}%
  \BibitemOpen
  \bibfield  {author} {\bibinfo {author} {\bibfnamefont {J.~H.}\ \bibnamefont
  {Eberly}},\ }\bibfield  {title} {\bibinfo {title} {Schmidt analysis of
  pure-state entanglement},\ }\href {https://doi.org/10.1134/S1054660X06060041}
  {\bibfield  {journal} {\bibinfo  {journal} {Laser Physics}\ }\textbf
  {\bibinfo {volume} {16}},\ \bibinfo {pages} {921} (\bibinfo {year}
  {2006})}\BibitemShut {NoStop}%
\bibitem [{\citenamefont {Sperling}\ and\ \citenamefont
  {Vogel}(2006)}]{Vogel2006}%
  \BibitemOpen
  \bibfield  {author} {\bibinfo {author} {\bibfnamefont {J.}~\bibnamefont
  {Sperling}}\ and\ \bibinfo {author} {\bibfnamefont {W.}~\bibnamefont
  {Vogel}},\ }\bibfield  {title} {\bibinfo {title} {The schmidt number as a
  universal entanglement measure},\ }\href
  {https://doi.org/https://doi.org/10.1088/0031-8949/83/04/045002} {\bibfield
  {journal} {\bibinfo  {journal} {Physica Scripta}\ }\textbf {\bibinfo {volume}
  {83}},\ \bibinfo {pages} {045002} (\bibinfo {year} {2006})}\BibitemShut
  {NoStop}%
\end{thebibliography}%

\end{document}